\documentclass[oldversion]{aa}

\usepackage[tbtags,fleqn]{amsmath}
\usepackage{amsfonts}
\usepackage{pstricks, pst-plot}
\usepackage{graphicx}
\usepackage{natbib}

\newcommand{\diff}{\mathrm d}

\newcommand{\mincir}{\raise
  -2.truept\hbox{\rlap{\hbox{$\sim$}}\raise5.truept \hbox{$<$}\ }}
\newcommand{\magcir}{\raise
  -2.truept\hbox{\rlap{\hbox{$\sim$}}\raise5.truept \hbox{$>$}\ }}


\begin{document}

\title{Hipparcos distance estimates of the Ophiuchus and the
  Lupus cloud complexes}
\titlerunning{Hipparcos distances of Ophiuchus and Lupus cloud complexes}
\author{Marco Lombardi\inst{1,2}, Charles J. Lada\inst{3}, and Jo\~ao
  Alves\inst{4}}
\authorrunning{M. Lombardi \textit{et al}.}
\offprints{M. Lombardi}
\mail{mlombard@eso.org}
\institute{%
  Space Telescope European Coordination Facility / ESO,
  Karl-Schwarzschild-Stra\ss e 2, 
  D-85748 Garching, Germany 
  \and 
  University of Milan, Department of Physics, via Celoria 16, I-20133
  Milan, Italy (on leave) 
  \and
  Harvard-Smithsonian Center for Astrophysics, Mail Stop 42, 60 Garden
  Street, Cambridge, MA 02138
  \and 
  Calar Alto Observatory -- Centro Astron\'omico Hispano Alem\'an,
  C/Jes\'us Durb\'an Rem\'on 2-2, 04004 Almeria, Spain}
\date{Received ***date***; Acceptet ***date***} 

\abstract{%
  We combine extinction maps from the Two Micron All Sky Survey
  (2MASS) with Hipparcos and Tycho parallaxes to obtain reliable and
  high-precision estimates of the distance to the Ophiuchus and Lupus
  dark complexes.  Our analysis, based on a rigorous
  maximum-likelihood approach, shows that the $\rho$-Ophiuchi cloud is
  located at $(119 \pm 6) \mbox{ pc}$ and the Lupus complex is located
  at $(155 \pm 8) \mbox{ pc}$; in addition, we are able to put
  constraints on the thickness of the clouds and on their orientation
  on the sky (both these effects are not included in the error
  estimate quoted above).  For Ophiuchus, we find some evidence that
  the streamers are closer to us than the core.  The method applied in
  this paper is currently limited to nearby molecular clouds, but it
  will find many natural applications in the GAIA-era, when it will be
  possible to pin down the distance and three-dimensional structure of
  virtually every molecular cloud in the Galaxy.  }

\maketitle

%

\section{Introduction}
\label{sec:introduction}

In recent years, much effort has been dedicated to the study of dark
molecular clouds and their dense cores.  One of the main motivations
for these investigations is the study of the process of star and
planet formation in its entirety, and a deeper understanding of the
effects of the local environment.  A key aspect of the scientific
analysis of a dark molecular cloud is its distance, which is related
to many physically relevant properties (in particular, the mass scales
as the square of the distance).  Unfortunately, the distance estimates
for many clouds are often plagued by very large uncertainties and it is
not rare to see in the literature measurements that differ by large
factors.  Often, this quantity is inferred by associating the cloud to
other astronomical objects whose distance is well known.  For example,
the Lupus complex is located near the Sco OB2 association
\citep{1978ApJS...38..309H, 1999AJ....117..354D}, whose distance is
estimated to be $\sim 150 \mbox{ pc}$ based on the photometry of OB
stars \citep{1991ESOSR..11....1R}.  However, this method is based on
some degree of arbitrariness when making the link between the cloud
and the other objects, and thus the deduced distance can be completely
unreliable.

The cloud complexes considered in this paper, the $\rho$ Ophiuchi and
the Lupus dark clouds, are good examples of how different distance
estimates made by different authors can be.  Quoted values for the
$\rho$-Ophiuchi cloud are in the range $120$--$165 \mbox{ pc}$, as
recently summarized by \citet{2004AJ....127.1029R}.
\citet{1981A&A....99..346C} estimated the distance to be $\sim 160
\mbox{ pc}$ from multi-color photometry of heavily absorbed stars in
the $\rho$ Oph core, while \citet{1998A&A...338..897K} suggested a
significantly smaller figure, $\sim 120 \mbox{ pc}$.  For Lupus, the
situation is even more extreme, with estimates from $100 \mbox{ pc}$
\citep{1998A&A...338..897K} to $190 \mbox{ pc}$
\citep{1998MNRAS.301L..39W}, both based on Hipparcos data; the most
widely accepted distance is in the middle of these two extremes, at
approximately $140 \mbox{ pc}$ \citep{1993AJ....105..571H,
  1999AJ....117..354D}.

These uncertainties severely hamper our understanding of the physical
properties of molecular clouds, and thus our knowledge of star
formation.  For example, an error by a factor two on the distance of a
cloud translates into an error by a factor four in the mass, and has a
thus a huge impact on the estimate of the density, size, stability,
and star formation efficiency of their cloud cores (see
\citealp{2007A&A...462L..17A} for further discussions on this point).

In this paper we present accurate distance measurements of the
Ophiuchus and Lupus complexes based upon Hipparcos and Tycho
parallaxes.  Our technique is statistically sound and, when applied to
nearby giant molecular clouds, is able to provide accurate distance
measurements and related errors.  The paper is organized as follows.
In Sect.~\ref{sec:basic-analysis} we present the main datasets used
and discuss a simple approach to the problem.  A more quantitative
technique is developed in Sect.~\ref{sec:likelihood-analysis}, where
we also presents the results obtained for the two clouds considered in
this paper.  The technique is further developed in
Sect.~\ref{sec:second-order-geom} to include the effects of the cloud
thickness and orientation.  Finally, in Sect.~\ref{sec:discussion} we
briefly discuss the results obtained.

\section{Basic analysis}
\label{sec:basic-analysis}

\begin{figure*}[!tbp]
  \begin{center}
    \includegraphics[width=\hsize]{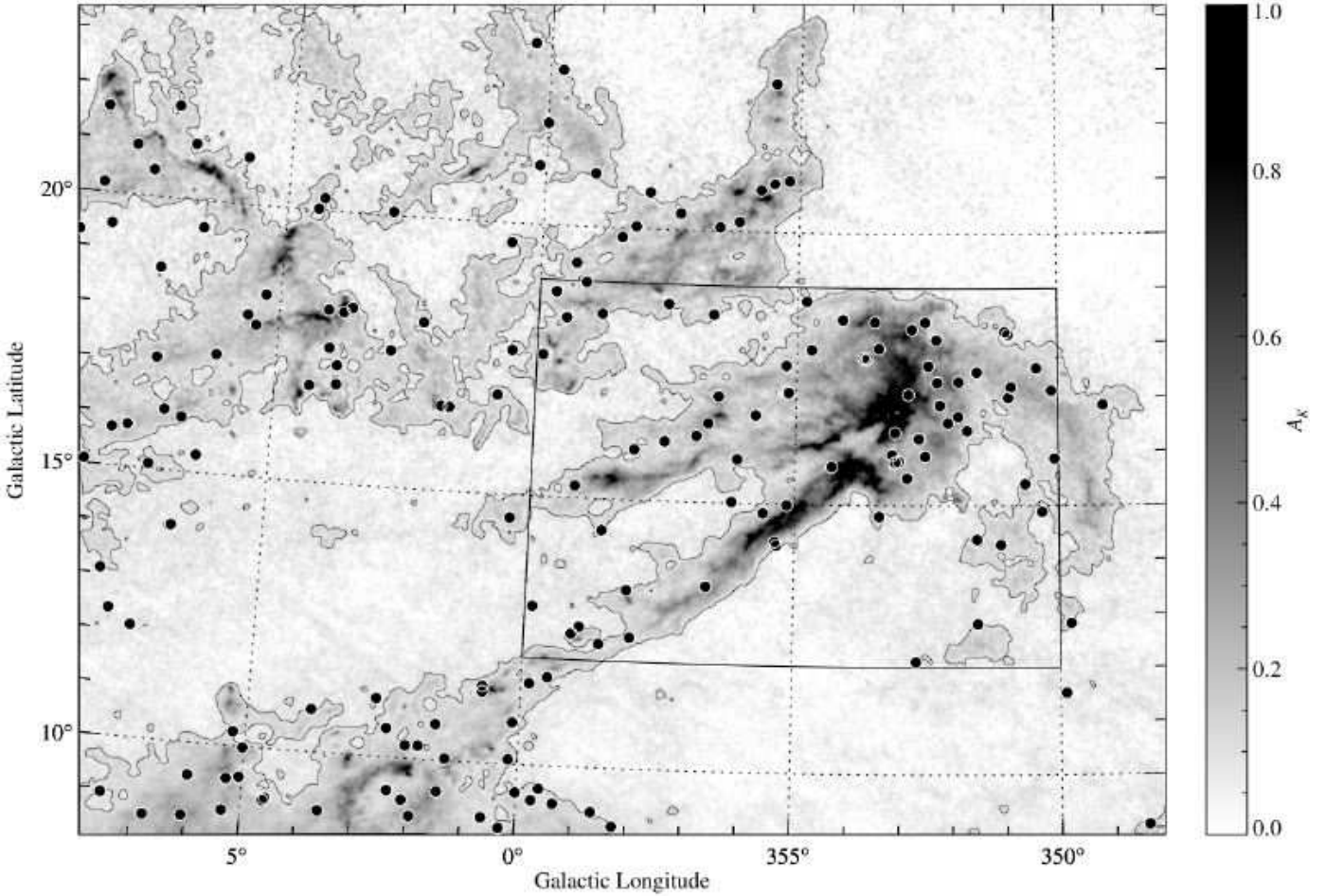}
    \caption{The Ophiuchus region considered.  The gray map shows
      2MASS/\textsc{Nicer} extinction, with overlaid the $A_K = 0.1
      \mbox{ mag}$ (smoothed) contour.  The dots indicate the position
      of the Hipparcos stars used for the analysis.  The boxed area
      indicates the central part of the cloud.}
    \label{fig:1}
  \end{center}
\end{figure*}

\begin{figure*}[!tbp]
  \begin{center}
    \includegraphics[width=\hsize]{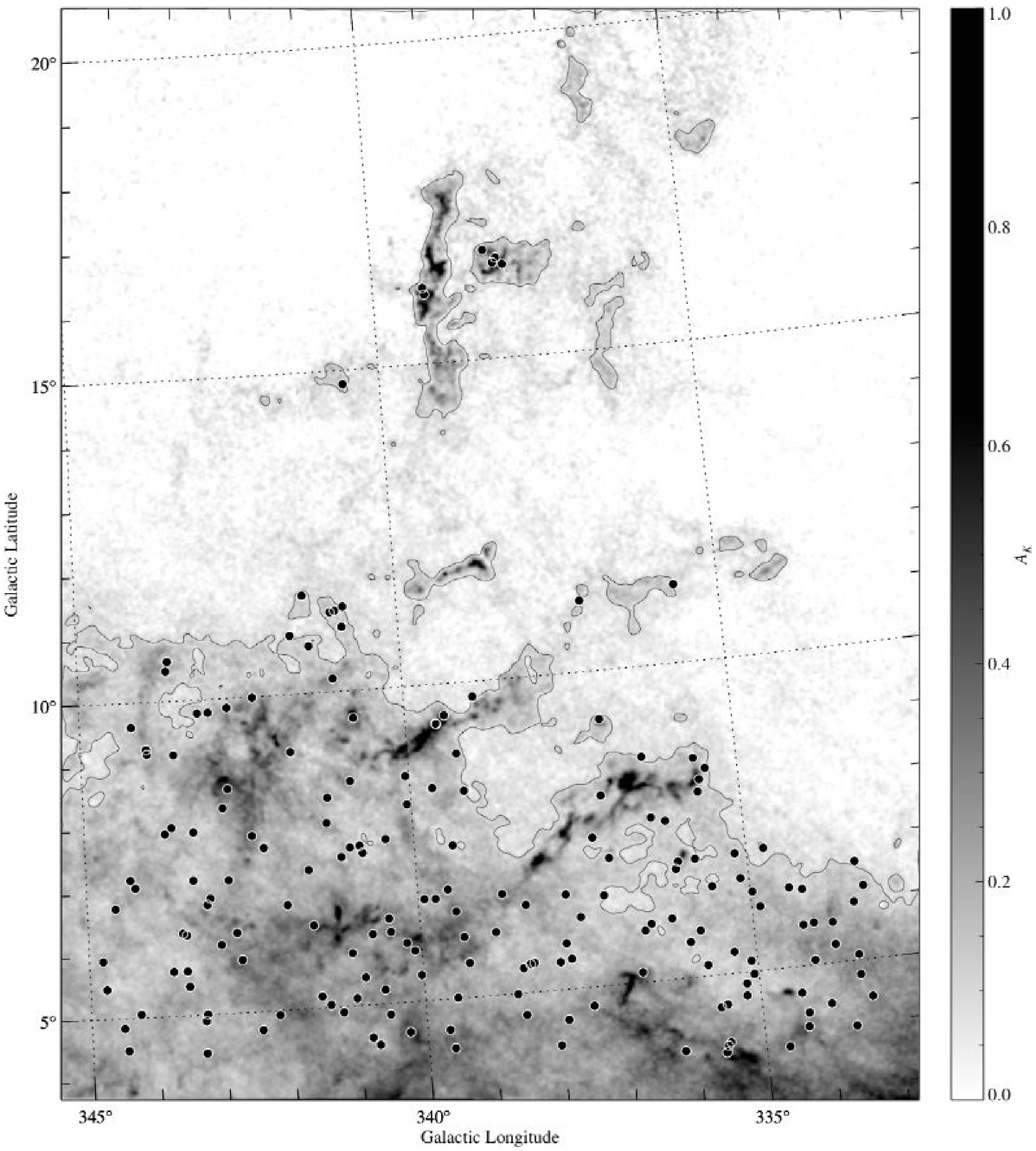}
    \caption{Same as Fig.~\ref{fig:1}, but for Lupus.}
    \label{fig:2}
  \end{center}
\end{figure*}

\begin{figure}[!tbp]
  \begin{center}
    \includegraphics[width=\hsize, bb=5 0 328 249]{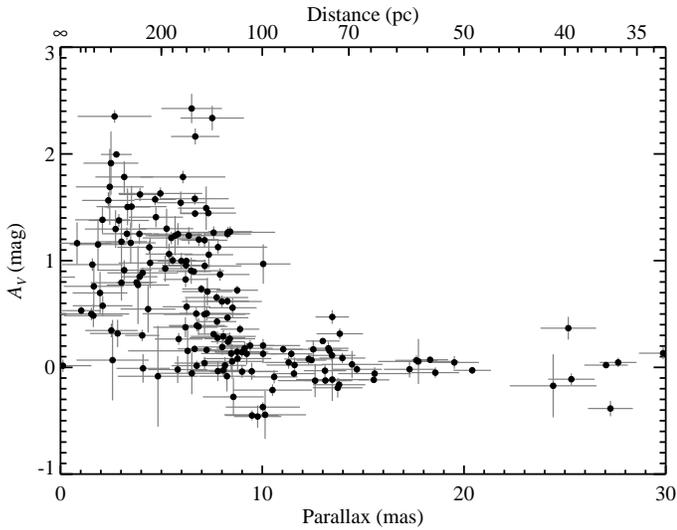}
    \caption{The reddening of Hipparcos and Tycho stars observed in
      direction of high integrated column densities ($A_K > 0.1 \mbox{
        mag}$) in the Ophiuchus field.  Only stars with relative error
      on the parallax smaller than $0.3$ were plotted here.  In this
      and other plots the reddening has been converted into $V$-band
      extinction by assuming a normal reddening law
      \citep{1985ApJ...288..618R}.}
    \label{fig:3}
  \end{center}
\end{figure}

\begin{figure}[!tbp]
  \begin{center}
    \includegraphics[width=\hsize, bb=5 0 331 231]{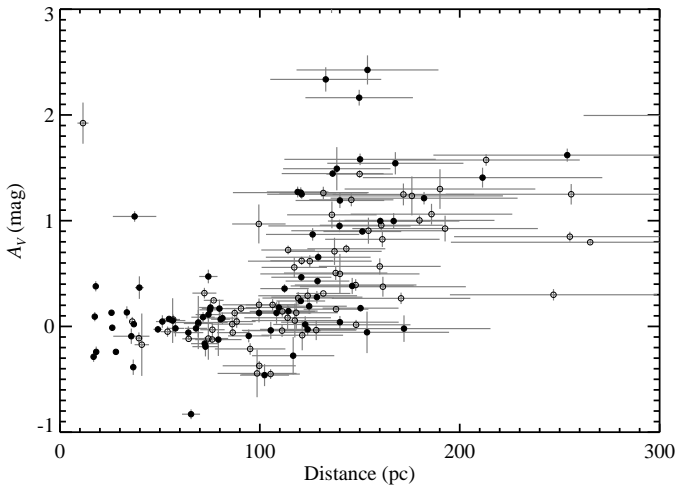}
    \caption{The reddening as a function of the Tycho distance for
      Ophiuchus stars.  The plot shows the same points as
      Fig.~\ref{fig:3}, but is done in distance instead of in
      parallaxes (and, as a result, error bars in distance are
      approximated).  Filled dots are stars for which the local
      extinction is $A_k > 0.15 \mbox{ mag}$, open dots stars with
      $A_K > 0.1 \mbox{ mag}$.  This figure can be better used to
      obtain a precise distance of the cloud, that we evaluate as $d =
      (120 \pm 10) \mbox{ pc}$.}
    \label{fig:4}
  \end{center}
\end{figure}

\begin{figure}[!tbp]
  \begin{center}
    \includegraphics[width=\hsize, bb=5 0 328 249]{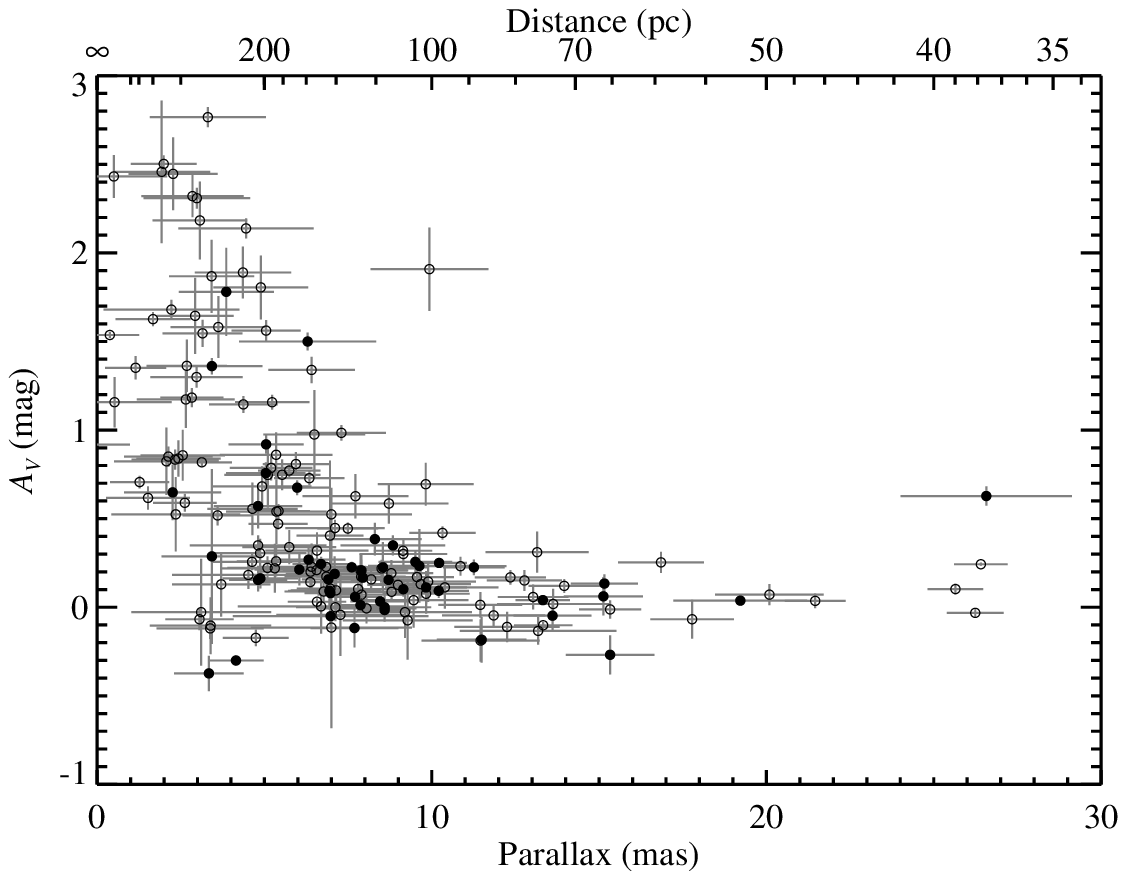}
    \caption{The reddening of stars as a function of their parallaxes
      in the Lupus region (filled circles: $A_V > 1.5 \mbox{ mag}$;
      open circles: $A_V > 1 \mbox{ mag}$).  Stars with relative
      parallax error larger than 0.5 were excluded from this plot.  If
      the cloud were a perfect ``wall'' of dust, we would obtain a
      Heaviside function with the discontinuity at the parallax of the
      cloud, much like Fig.~\ref{fig:3} for Ophiuchus.  Measurement
      errors, the complex structure of the cloud, and the small number
      statistics all make this plot to more scattered and thus less
      clear.}
    \label{fig:5}
  \end{center}
\end{figure}

Our technique is based on a simple fact: stars observed through high
column-densities must exhibit a significant reddening, while
foreground stars will show no reddening even when observed in
projection toward dense regions of the cloud.  The first step was thus
to make reliable extinction maps of the $\rho$-Ophiuchi and Lupus
clouds in order to delineate the cloud boundaries and regions to be
considered for the distance analysis.  For the purpose, we used
approximately $42$ million stars from the Two Micron All Sky Survey
(2MASS) point source catalog to construct a $1\,672$ square degrees
\textsc{Nicer} \citep{2001A&A...377.1023L} extinction map of the
Ophiuchus and Lupus dark nebul\ae.  The map, described elsewhere
\citep{Lombardi07a}, a has a resolution of $3\mbox{ arcmin}$ and a
$3\sigma$ detection level of $0.5$ visual magnitudes.  We considered
the two regions shown in Figs.~\ref{fig:1} and \ref{fig:2},
approximately corresponding to the galactic coordinates
\begin{align}
  \label{eq:1}
  -12^\circ < l & {} < 2^\circ &
  9^\circ < b & {} < 24^\circ \\
  \text{for Ophiuchus, and} \notag\\
  \label{eq:2}
  353^\circ < l & {} < 345^\circ &
  4^\circ < b & {} < 19^\circ
\end{align}
for Lupus, and selected there all Hipparcos and Tycho
\citep{1997A&A...323L..49P} parallax measurements of stars in regions
characterized by a relatively high column densities (more precisely,
in regions with $A_K > 0.1 \mbox{ mag}$ in our 2MASS extinction map).
Finally, in order to estimate the \textit{intrinsic\/} colors of the
Hipparcos stars, we obtained their spectral types.  For the purpose,
we used the All-Sky Compiled Catalogue (ASCC-2.5;
\citealp{2001KFNT...17..409K}), which lists $2\,501\,304$ stars taken
from several sources (mainly the Hipparcos-Tycho family, the
ground-based Position and Proper Motion family, and the Carlsberg
Meridian Catalogs), and the ``Tycho-2 Spectra Type Catalog''
(Tycho-2spec, III/231; \citealp{2003AJ....125..359W}), which
cross-references $351\,863$ Tycho stars with spectral catalogues
(mainly the Michigan catalogs).  In summary, our joined catalog
contains for each star observed in projection with relatively large
extinction regions the parallax and its related error, the spectral
type, and a set of visual magnitudes with errors.

We then considered all Hipparcos stars with well determined parallaxes
(we required the parallax error to be smaller than $3 \mbox{ mas}$) in
the region considered, and estimated the intrinsic $B - V$ of each
star from its spectral type (for this purpose, we used the color tables
in \citealp{LB}, p.~15, and the normal reddening law of
\citealp{1985ApJ...288..618R}).  Finally, we evaluated the extinction
of each star by comparing its observed $B-V$ color with its estimated
intrinsic color.  As final selection, we required stars to have a
$V$-band extinction error smaller than $1 \mbox{ mag}$, which left 180
objects in the Ophiuchus region, and 186 objects in the Lupus region.
A plot of the star column density versus the Hipparcos parallax for
both cloud complexes is shown in Figs.~\ref{fig:3} and \ref{fig:5}; a
distance-reddening zoom plot for Ophiuchus is presented in
Fig.~\ref{fig:4}.

Figure~\ref{fig:3} proves the effectiveness of the parallax method:
from the right to the left, stars are initially found close to the
$A_V = 0 \mbox{ mag}$ line, with a relatively low dispersion; then, as
we go to smaller parallaxes, an impressive ``wall'' of stars with
significant reddening is observed (approximately at $\pi = 8 \mbox{
  mas}$).  Since for Fig.~\ref{fig:3} we only selected stars observed
against high-column density regions in Ophiuchus, and we further
excluded stars with suboptimal measurement errors, the
parallax-reddening plot appears extremely clear and gives alone an
accurate measurements of the distance of the cloud, that we estimate
to be approximately $120 \mbox{ pc}$.

The same plot for the Lupus region, shown in Fig.~\ref{fig:5}, appears
to be slightly less clear, for several reasons.  First, Lupus has a
complex structure and is composed of several subclouds spanning
several degrees (see Fig.~\ref{fig:2}); hence, it is not unlikely
that different clouds are located at different distances (see
discussion below).  In addition, the Lupus cloud complex seems to be
at a larger distance than Ophiuchus, at the limit of the Hipparcos
sensitivity, and hence the plot of Fig.~\ref{fig:5} is perhaps not as
crystal clear as the one of Fig.~\ref{fig:3}.  Still, a preliminary
qualitative analysis indicates a cloud distance close to $150 \mbox{
  pc}$ for Lupus.

An intrinsic problem of this technique is that, in this basic
formulation, it does not allow a simple, robust estimate of the error
associated with the distance measurement.  In addition, the value
provided must be regarded as an \textit{upper limit},
since the distance is typically estimated from the first star that
shows a significant reddening.

\section{Likelihood analysis}
\label{sec:likelihood-analysis}

\begin{figure}[!tbp]
  \begin{center}
    \includegraphics[bb=0 0 328 249, width=\hsize]{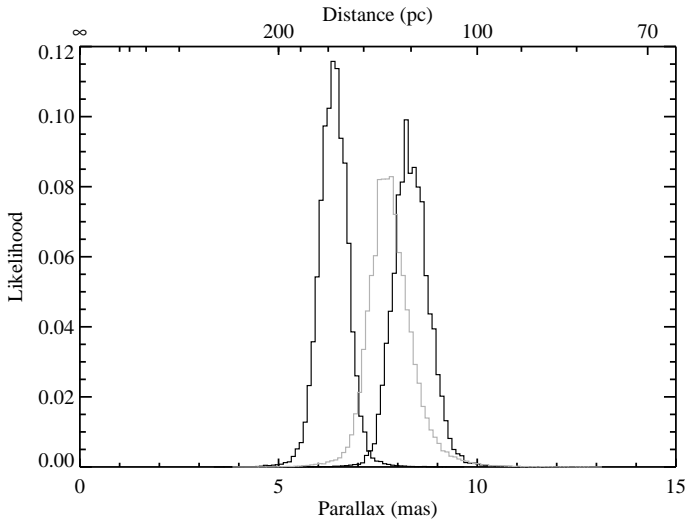}
    \caption{The likelihood function of Eq.~\eqref{eq:6} for Lupus
      (left black histogram) and Ophiuchus (right black histogram) as a
      function of the cloud parallax $\pi_\mathrm{cloud}$ marginalized
      over all the other parameters.  Both histograms shows marked
      peaks at $\pi_\mathrm{Lup} \simeq 7 \mbox{ mas}$ and
      $\pi_\mathrm{Oph} \simeq 9 \mbox{ mas}$, corresponding to
      distances of approximately $119 \mbox{ pc}$ and $155 \mbox{
        pc}$, respectively.  The grey histogram refers to the
      likelihood function for the central region of Ophiuchus (marked
      window in Fig.~\ref{fig:1}).}
    \label{fig:6}
  \end{center}
\end{figure}

A less qualitative estimate can be obtained by using a statistical
model for the parallax-reddening relation.  Following
\citet{2006A&A...454..781L}, we assumed that the $A_V$ measurement for
foreground stars is small, while a fraction $f$ of the background
stars is distributed as a normal variable with relative large mean and
variance.  Mathematically, we considered the two distributions
\begin{align}
  \label{eq:3}
  p^\mathrm{fg}(A_V) = {} & \mathrm{Gau}(A_V | A_V^\mathrm{fg}, 
  \sigma_{A_V}^{\mathrm{fg}2} + \sigma_{A_V}^2) \; , \\
  \label{eq:4}
  p^\mathrm{bg}(A_V) = {} & \mathrm{Gau}(A_V | A_V^\mathrm{bg}, 
  \sigma_{A_V}^{\mathrm{bg}2} + \sigma_{A_V}^2) \; ,
\end{align}
where we denoted with $\mathrm{Gau}(x | \bar{x}, \sigma_x^2)$ the
value at $x$ of a normal probability density with mean $\bar{x}$ and
variance $\sigma^2_x$.  The two distributions $p^\mathrm{fg}$ and
$p^\mathrm{bg}$ refers to foreground (i.e., almost unextincted) and
background (heavily extincted) stars.  These are taken to be normal
distributions with variances given by the sum of the intrinsic scatter
of column densities ($\sigma_{A_V}^{\mathrm{fg}2}$ and
$\sigma_{A_V}^{\mathrm{bg}2}$) and of the measurement error for the
star considered ($\sigma_{A_V}^2$).  Note, in particular, that
$\sigma_{A_V}^\mathrm{bg}$ accounts for the local changes in the cloud
column density with respect to the average value $A_V^\mathrm{bg}$,
while $\sigma_{A_V}^\mathrm{fg}$ is introduced to allow for
underestimates of the photometric errors on the star magnitudes or for
foreground, thin nearby clouds.  Finally, the probability distribution
$p(A_V | \pi)$ for the $A_V$ measurement for a star at parallax $\pi$
was taken to be
\begin{equation}
  \label{eq:5}
  p(A_V | \pi) = 
  \begin{cases}
    p^\mathrm{fg}(A_V) & 
    \text{if $\pi > \pi_\mathrm{cloud} \; ,$} \\[0.5em]
    (1 - f) p^\mathrm{fg}(A_V) + f p^\mathrm{bg}(A_V) & 
    \text{if $\pi \le \pi_\mathrm{cloud} \; .$}
  \end{cases}
\end{equation}
Note that the parameter $f \in [0, 1]$ denotes the ``filling factor,''
i.e.\ the probability that a background star is extincted by the cloud.

We considered this simple model leaving all parameters $\bigl\{
\pi_\mathrm{cloud}, f, A_V^\mathrm{fg}, \sigma_{A_V}^\mathrm{fg},
A_V^\mathrm{bg}, \sigma_{A_V}^\mathrm{bg} \bigr\}$ free.  In order
to assess the goodness of a model, we computed the likelihood
function, defined as
\begin{multline}
  \label{eq:6}
  \mathcal{L}(\pi_\mathrm{cloud}, f, A_V^\mathrm{fg}, \sigma_{A_V}^\mathrm{fg},
  A_V^\mathrm{bg}, \sigma_{A_V}^\mathrm{bg}) = {} \\ 
  \prod_{n=1}^N \int_0^\infty p\bigl( A_V^{(n)} \bigm| \pi^{(n)} 
  \bigr) p\bigl( \pi^{(n)} \bigm| \hat\pi^{(n)} \bigr) \, \diff \pi \; ,
\end{multline}
where the product is carried over all $N$ stars of Figs.~\ref{fig:3}
and \ref{fig:5}, and where $p\bigl( \pi^{(n)} \bigm| \hat\pi^{(n)}
\bigr)$ is the probability distribution that the $n$-th star with
measured parallax $\hat\pi^{(n)}$ has a true parallax $\pi^{(n)}$.
Note that the integral appearing in Eq.~\eqref{eq:6} properly takes
into account the possibility that a star with a measured parallax
$\hat\pi^{(n)} < \pi_\mathrm{cloud}$ (or $\hat\pi^{(n)} >
\pi_\mathrm{cloud}$) is incorrectly taken as background (respectively,
foreground).  For this probability we used a normal distribution:
\begin{equation}
  \label{eq:7}
  p\bigl( \hat\pi^{(n)} \bigm| \pi^{(n)} \bigr) = \mathrm{Gau}\bigl(
  \hat\pi^{(n)} \bigm| \pi^{(n)}, \hat\sigma_{\pi}^{(n)2} \bigr) \; ,
\end{equation}
where $\hat\sigma_{\pi}^{(n)}$ is the estimated error on the parallax.
Note that the use of normal distributions in Eqs.~\eqref{eq:3},
\eqref{eq:4}, and \eqref{eq:7} makes it possible to perform the
integral \eqref{eq:6} analytically in terms of the error function erf.

The likelihood function \eqref{eq:6} was analysed using Monte Carlo
Markov Chains (MCMC; see, e.g. \citealp{Tanner}).  Figure~\ref{fig:6}
shows the likelihood as a function of the cloud distance, marginalized
with respect to all other parameters.  From this figure we can
trivially evaluate the confidence regions for the estimated distances;
in particular, we obtain $d_\mathrm{Oph} = (119 \pm 6) \mbox{ pc}$,
and $d_\mathrm{Lup} = (155 \pm 8) \mbox{ pc}$, both at the $68\%$
confidence level (since the likelihood function is well approximated
by a Gaussian, other confidence bounds can easily derived from the
errors quoted here).  We stress that the errors quoted here include
only statistical errors, as deduced from the parallaxes and reddening
uncertainties; other effects, such as the thickness of the cloud or
its orientation in the sky are not taken into account and are
discussed below in Sect.~\ref{sec:second-order-geom}.  We stress also
that the Lupus distance estimate approximately refers to the average
location of background stars in the field (i.e., to $l = 339^\circ$
and $b = 7^\circ$), approximately $5.5$ degrees south than the center
of the field considered, and that only a few stars are observed
through Lupus~1.  This difference can play a significant role in case
of noticeable tilts in the cloud orientation or in case of multiple
components at different distances (see below).

Note that the reliability of the maximum-likelihood technique
presented here was also checked with numerical simulations.  It is
well known that the maximum-likelihood method is, under certain
circumstances (verified in our case), asymptotically unbiased;
however, we decided to test its behaviour in our specific case, with a
limited (although large) number of data.  For the purpose, we simulated
a ``thin'' cloud together with observation of a population of stars.
We included reasonable errors on both the star parallaxes and column
densities, and recovered the distance of the cloud using the same
method described here.  These simulations confirmed that our estimate
is essentially \textit{unbiased\/} (i.e., that on average it returns
the true distance of the cloud), and that the error provided is
reliable.  

We also considered a two-screen configuration, with two partially
overlapping clouds at different distances, and checked the distance
provided by the method.  Interestingly, the simulation showed that in
this case the method tends to return a ``weighted'' distance, i.e.\ a
weighted mean of distance of the two screens (where the weights are
essentially proportional to the number of stars intercepted by the two
screens and by the column densities associated with the two clouds).
This result is very encouraging, since it guarantees that, at least in
the relatively simple cases considered, the maximum likelihood provide
a sensible and unbiased estimate of the cloud distance.  Note,
however, that the last result also shows that in case of composite
clouds with significantly different densities of background stars
(e.g., because of their different galactic latitude), the
maximum-likelihood technique will be biased toward the distance of the
cloud with the highest density of stars.  This point is particularly
relevant for Lupus, a cloud complex that spans several degrees in
galactic latitude, and for which there are indications of possible
different distances among the various subclouds (see below
Sect.~\ref{sec:cloud-orient-tilt}; cf.\ also
\citealp{2006MNRAS.366..238A}).

\section{Second order geometrical effects}
\label{sec:second-order-geom}

Given the angular size of the regions considered, it is not unlikely
that the cloud complexes have a thickness of $15 \mbox{ pc}$ or more.
Similarly, even for ``thin'' clouds, we can easily imagine that they
are not perfectly perpendicular to the line of sight and that
different regions of them are at different distances.  All these
geometrical effects can be introduced in the maximum-likelihood
analysis discussed above.

\subsection{Cloud thickness}
\label{sec:cloud-thickness}

The thickness of a molecular cloud can be investigated by modifying
the probability distribution of Eq.~\eqref{eq:5}.  This can be done in
several ways, which corresponds to different ``definitions'' of the
thickness of a cloud.  We chose to parametrize the thickness as
a continuous, linear transition in the filling factor $f$, which
effectively becomes a function of the parallax.  Hence, we
wrote
\begin{equation}
  \label{eq:8}
  p(A_V | \pi) = f(\pi) p^\mathrm{bg}(A_V) + \bigl[ 1 - f(\pi) \bigr]
  p^\mathrm{fg}(A_V) \; ,
\end{equation}
where
\begin{equation}
  \label{eq:9}
  f(\pi) = 
  \begin{cases}
    0 & \text{if $\pi > \pi_1 \; ,$} \\[0.5em]
    f_\mathrm{max} (\pi - \pi_1) / (\pi_2 - \pi_1) & 
    \text{if $\pi_2 < \pi < \pi_1 \; .$} \\[0.5em]
    f_\mathrm{max} & \text{if $\pi < \pi_2 \; .$}
  \end{cases}
\end{equation}
In other words, stars up to the lower distance of the cloud ($\pi >
\pi_1$) show no extinction, and thus their column density is
distributed according to $p^\mathrm{fg}(A_V)$; similarly, stars
background to the cloud ($\pi < \pi_2$) follow, with probability
$f_\mathrm{max}$ (the usual filling factor), a distribution
$p^\mathrm{bg}(A_V)$; finally, stars embedded in the cloud ($\pi_2 <
\pi < \pi_1$) present an effective filling factors that raises
linearly within the cloud from $0$ to $f_\mathrm{max}$.  Clearly,
both parallaxes $\pi_1$ and $\pi_2$ were taken as free parameters.
Note that, similarly to Eq.~\eqref{eq:5}, the use of a simple linear
function for $f(\pi)$ guarantees that we can express analytically the
integral of Eq.~\eqref{eq:6}.

\begin{figure}[!tbp]
  \begin{center}
    \includegraphics[bb=01 0 324 243, width=\hsize]{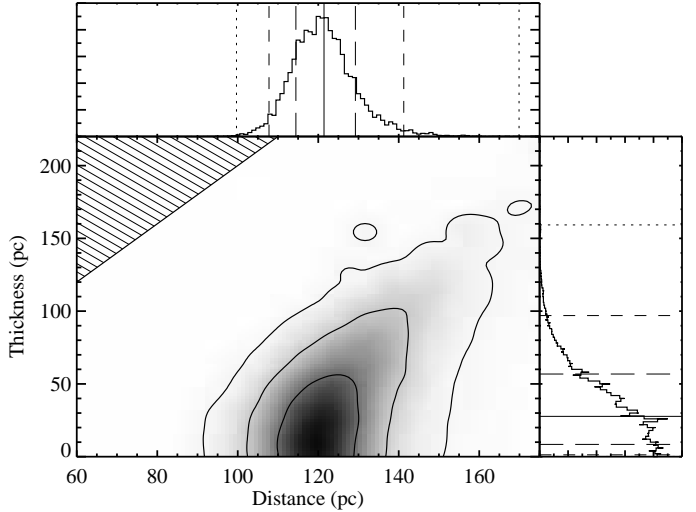}
    \caption{A density plot of the likelihood of Eq.~\eqref{eq:6} with for
      Ophiuchus with the extinction probability distribution
      \eqref{eq:8}, as a function of the cloud distance and thickness
      (the likelihood has been marginalized with respect to all other
      parameters).  The three curves enclose regions of $68\%$,
      $95\%$, and $99.7\%$ confidence level, while the line-filled
      corner in the top-left indicate a region that is physically
      excluded (observer embedded in the cloud).  The two plots at the
      top and at the right show the distance and thickness
      marginalized probabilities, together with the median value
      (solid line) and the usual three confidence regions (dashed and
      dotted lines).}
    \label{fig:7}
  \end{center}
\end{figure}

\begin{figure}[!tbp]
  \begin{center}
    \includegraphics[bb=01 0 324 243, width=\hsize]{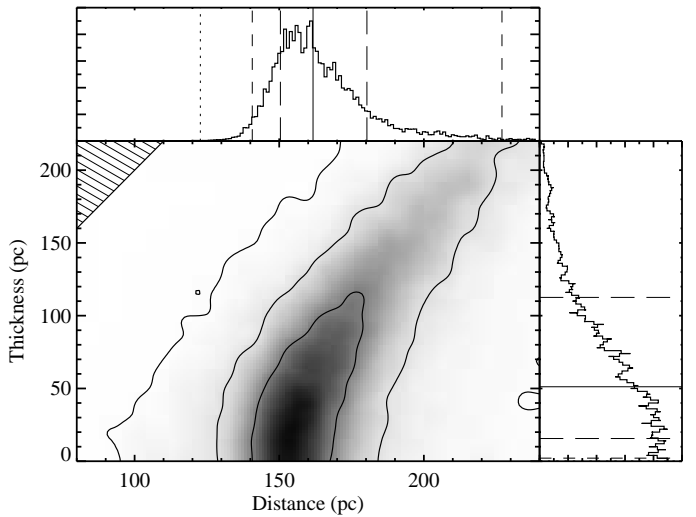}
    \caption{Same as Fig.~\ref{fig:7} for Lupus.}
    \label{fig:8}
  \end{center}
\end{figure}

The results of a MCMC analysis of the likelihood function \eqref{eq:6}
with the distribution of Eq.~\eqref{eq:8} are shown in
Fig.~\ref{fig:7} for Ophiuchus and in Fig.~\ref{fig:8} for Lupus.  The
plots in these figures indicate that, unfortunately, the quality and
quantity of the available parallaxes is not sufficient to put strong
constraints on the thickness of the clouds.  In particular, for
Ophiuchus we obtained a median thickness of $28^{+29}_{-19} \mbox{
  pc}$, but the errors are clearly very large; the situation is even
more extreme for Lupus, with a thickness estimate of $51^{+61}_{-35}
\mbox{ pc}$.  Still, this analysis is important because it can in
principle provide different distributions for the distances of the two
clouds (see top plots in Figs.~\ref{fig:7} and \ref{fig:8}) with
respect to the ones obtained from the simpler analysis of
Sect.~\ref{sec:likelihood-analysis} (Fig.~\ref{fig:6}).  In this
respect, the fact that the two results obtained are almost identical
confirms the robustness of our distance estimates.  In addition, the
much larger thickness estimate for Lupus and the flat plateau evident
in the right plot of Fig.~\ref{fig:8} suggest that the apparent
thickness of Lupus might be the result of different Lupus subclouds
being at different distances (see below
Sect.~\ref{sec:cloud-orient-tilt}).  Finally, we note that the method
proposed here to study the cloud thickness will certainly provide much
more interesting results when a large statistics of high-quality
parallaxes will become available.

\subsection{Cloud orientation (tilt)}
\label{sec:cloud-orient-tilt}

In order to investigate the presence of possible tilts in the
orientation of molecular clouds, we considered a simple model of a
thin cloud located at a distance that is a linear function of the sky
coordinates:
\begin{equation}
  \label{eq:10}
  d(x, y) = d_0 + d_x x + d_y y \; .
\end{equation}
In this equation, $d_0$ is the distance of the cloud at a reference
point (typically, the center of the region considered), and the sky
coordinates $(x,y)$ are taken to be measured in pc and oriented along
the horizontal and vertical axes of Figs.~\ref{fig:1} and \ref{fig:2}
(so that $d_x$ and $d_y$ are pure numbers).  This simple prescription
can be directly introduced in Eq.~\eqref{eq:5} without any further
significant modification of the model.

\begin{figure}[!tbp]
  \begin{center}
    \includegraphics[bb=5 7 324 324, width=\hsize]{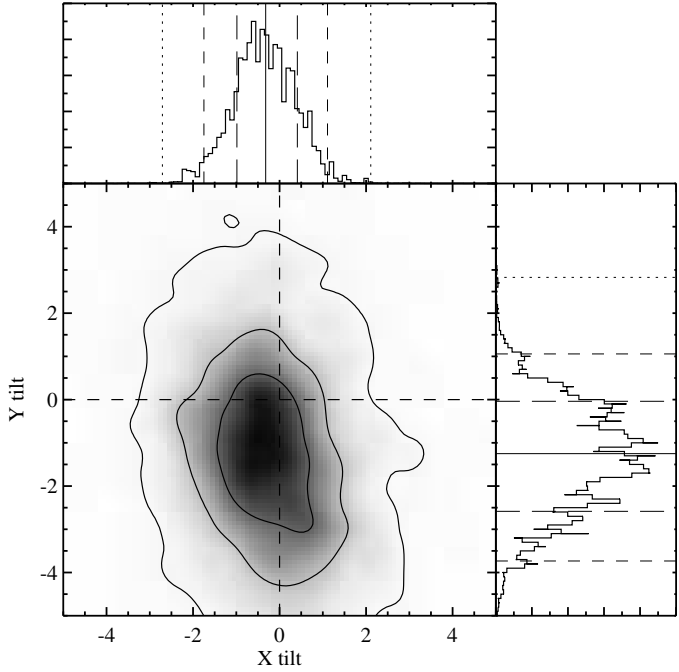}
    \caption{A density plot of the derived probability distribution
      for the tilt parameters $d_x$ and $d_y$ of Eq.~\eqref{eq:10} in
      Ophiuchus.  Similarly to Fig.~\ref{fig:7}, the three curves
      enclose regions of $68\%$, $95\%$, and $99.7\%$ confidence
      level and the two plots at the top and at the right show the
      X tilt and Y tilt marginalized probabilities.}
    \label{fig:9}
  \end{center}
\end{figure}

\begin{figure}[!tbp]
  \begin{center}
    \includegraphics[bb=5 7 324 324, width=\hsize]{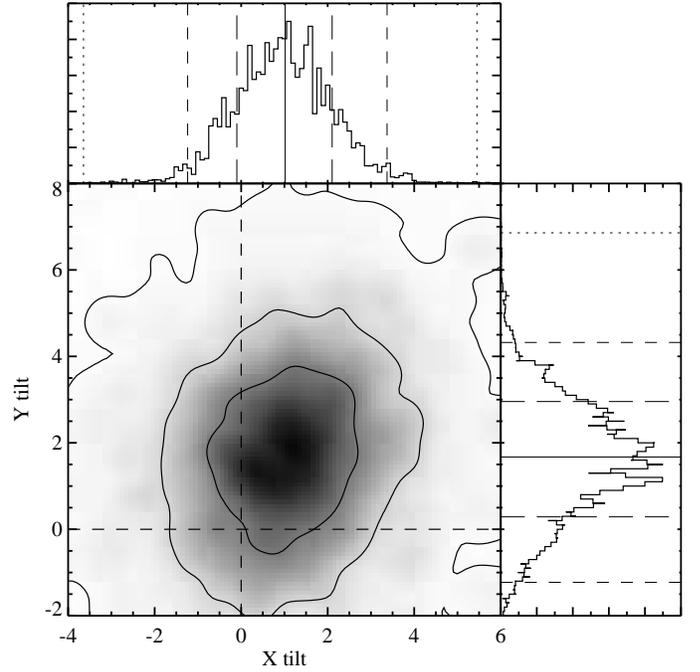}
    \caption{Same as Fig.~\ref{fig:9} for Lupus.}
    \label{fig:10}
  \end{center}
\end{figure}

We studied again this model with MCMC and obtained the probability
distribution for $(d_x, d_y)$ shown in Figs.~\ref{fig:9} and
\ref{fig:10}.  In summary, for Ophiuchus we marginally detected a
small tilt oriented such that the Ophiuchus streamers would be closer
to us than the $\rho$ Ophiuchi core.  Indeed, if we repeat the simple
analysis of Sect.~\ref{sec:likelihood-analysis} to the window marked
with a solid line in Fig.~\ref{fig:1}, the denser central regions of
Ophiuchus that include the well known $\rho$ Oph core, the distance
estimate increases to $(128 \pm 8) \mbox{ pc}$.  In contrast, for
Lupus we detect a vertical gradient in the distance, with the
high-galactic latitude regions at larger distances than the
low-galactic latitude ones (note that, according to the right plot of
Fig.~\ref{fig:10}, the gradient in galactic latitude is detected at a
$3 \sigma$ level).  Hence, this result together with the large extent
of Lupus on the sky seems to indicate that this complex might not be
composed of physically related clouds, as also proposed by
\citet{1999A&A...352..574B} and \citet{2001A&A...373..714K}.
Unfortunately, in the current form the method presented here cannot
\textit{automatically\/} distinguish multiple components of a cloud
complex located at different distances.  Of course, it is always
possible to repeat the analysis in different subclouds, but this
requires an a priori assumption on the subcloud division.  For Lupus
subclouds at high galactic latitudes, in particular, this is manifestly
not viable given the relatively small number of stars present there.
As described above, we applied this technique to the central region of
the Ophiuchus complex, obtaining a somewhat higher distance for it.
We also considered subregions of Ophiuchus occupying different sectors
of Fig.~\ref{fig:1} (defined by the two diagonals, the central
vertical line, and the central horizontal line).  However, within the
relatively large errors of these analyses, we could not identify any
significant difference in the distance.  Note that this result is
consistent with large errors shown by the plot of Fig.~\ref{fig:9},
which shows that a ``flat'' screen molecular cloud is still consistent
with our data.

\subsection{Effects of diffuse ISM}
\label{sec:effects-diffuse-ism}

So far, in our analysis we have implicitly assumed that the only
significant extinction is associated with the molecular cloud.  In
reality, the dust present in the interstellar medium (ISM) is known to
produce an average $V$-band extinction of approximately $1 \mbox{ mag
  kpc}^{-1}$ along the galactic plane.  In principle, it is not
difficult to take into account this effect by replacing
Eqs.~\eqref{eq:3} and \eqref{eq:4} with
\begin{align}
  \label{eq:11}
  p^\mathrm{fg}(A_V | \pi) = {} & \mathrm{Gau}(A_V | A_V^\mathrm{fg} +
  \alpha / \pi, \sigma_{A_V}^{\mathrm{fg}2} + \sigma_{A_V}^2) \; , \\
  \label{eq:12}
  p^\mathrm{bg}(A_V) = {} & \mathrm{Gau}(A_V | A_V^\mathrm{bg} +
  \alpha / \pi, 
  \sigma_{A_V}^{\mathrm{bg}2} + \sigma_{A_V}^2) \; ,
\end{align}
where $\alpha \simeq 1 \mbox{ mag mas}$.  In practice, however, given
the relatively large scatter $\sigma_{A_V}^\mathrm{fg}$ of
$A_V^\mathrm{fg}$ compared to $\alpha / \pi$ for the cloud considered
in this paper, the effect of the ISM can be safely neglected for the
foreground star probability distribution $p^\mathrm{fg}$.  In reality,
we argue that the same conclusion applies also to the background
distribution $p^\mathrm{bg}$.  First, we note that the large majority
of Hipparcos stars have relatively large parallaxes: for example, in
the Ophiucus region we found only 10 objects out of 180 with
parallaxes larger than $2 \mbox{ mas}$.  In addition, the combined
effect of a large scatter in $\sigma_{A_V}^\mathrm{bg}$ of
$A_V^\mathrm{bg}$ and of the filling factor $f$ makes the detection of
a gradient in $A^\mathrm{bg}$ currently unrealistic.  A specific
maximum-likelihood analysis using Eqs.~\eqref{eq:11} and \eqref{eq:12}
has been carried out, and the results show that the arguments just
presented hold.  In particular, no significant differences in the
distance determinations were found using either a fixed $1 \mbox{ mag
  mas}$ or a variable value for $\alpha$.

We stress, however, that with the advent of GAIA, it will be possible
to apply the method described in this paper to significantly more
distant clouds, for which the cumulative effect of the diffuse ISM can
play a relevant role.

\section{Discussion}
\label{sec:discussion}

Our analysis indicates that the Ophiuchus cloud is very likely to be
closer than the we think: the ``standard'' distance, $160 \mbox{ pc}$,
is indeed excluded at $5 \sigma$.  Not surprisingly, our estimate is
in excellent agreement with the one obtained by
\citet{1998A&A...338..897K} using a similar technique (but note that
the two results are based on a different datasets and use different
statistical methods).  

Recently, \citet{WilsonPhD} and \citet{2007arXiv0709.0505M} reported
an Hipparcos estimated distance of Ophiuchus of $d = 136^{+8}_{-7}
\mbox{ pc}$ and $d = 135^{+8}_{-7} \mbox{ pc}$, respectively, results
which are marginally in agreement with our distance estimate of the
Ophiuchus core.  \citeauthor{2007arXiv0709.0505M} mainly focused on
the Lynds~168 cloud, and inferred the cloud distance from the
parallaxes of seven stars presumably associated with illuminated
nebulae.  Although his method is plagued by a potential uncertainty,
the real association between the Hipparcos stars and the Ophiuchus
cloud complex, and suffers from the small number statistics (7
parallaxes), the agreement obtained is reassuring and suggests that we
finally have in hand a reliable estimate of the Ophiuchus cloud
complex.  \citet{WilsonPhD}, instead, used a more rigorous
maximum-likelihood approach, but his analysis is substantially
different from ours.  In particular, his method is based on a
preliminary classification of stars as foreground and background based
on their observed reddening; the distance of the cloud is then
inferred from the constraints imposed by the parallaxes of the stars
considered.  In contrast, in our method we do not need to explicitly
classify stars, but on the contrary can use directly all data in the
likelihood (an advantage of this is that we are able to use in our
method also stars that cannot be uniquely classified as foreground or
background because, e.g., of their large photometric errors).

Very recently, \citet{Loinard} used phase-referenced multi-epoch Very
Long Baseline Array (VLBA) observations to measure the trigonometric
parallax of the Ophiucus star-forming region.  Their method is based
on accurate (to better than a tenth of a milli-arcsecond), absolute
measurements of the position of individual magnetically active young
stars though to be associated with the star-forming region.  If
multi-epoch data are available, one can infer the parallax of the
star, and thus of the star-forming regions, with exquisite accuracy
(below $1 \mbox{ pc}$ for objects within $\sim 200 \mbox{ pc}$).  This
technique was applied to four stars in the $\rho$-Ophiuchi region, and
interestingly it lead to somewhat contradictory results: two objects
associated with the sub-condensation Oph~A (S1 and DoAr21) have a
measured parallax close to $\sim 120 \mbox{ pc}$, and thus in
excellent agreement with our estimate; two others, associated with the
condensation Oph~B (VSSG14 and VL5), have a much higher distance of
$\sim 165 \mbox{ pc}$.  A possible explanation of this is provided by
a closer examination of the two sources in the Oph~B condensation.
From their spectral energy distributions (SEDs), they appear to be
Class~III A-type stars, with little or no circumstellar emitting dust, and
an extinction as large as $50 \mbox{ mag}$ in the $V$ band (see
\citealp{1989ApJ...340..823W}, and especially
\citealp{1994ApJ...434..614G} for a discussion on the individual SEDs
of VSSG14 and VL5).  These results indicate that VSSG14 and VL5 are
probably background stars, not directly associated with the Ophiucus
complex, and likely to be part of a smaller OB association at $\sim
165 \mbox{ pc}$.

The distance of the Lupus complex is substantially in agreement with
the ones reported in the literature, but both the thickness analysis
and the orientation analysis suggest noticeable differences in the
various Lupus subclouds.  

\acknowledgements 

We thank Tom Dame and Alex Wilson for useful discussions, and the
referee, Jens Knude, for helping us improve the paper with many useful
comments.  This research has made use of the 2MASS archive, provided
by NASA/IPAC Infrared Science Archive, which is operated by the Jet
Propulsion Laboratory, California Institute of Technology, under
contract with the National Aeronautics and Space Administration.  This
paper also made use of the Hipparcos and Tycho Catalogs (ESA SP-1200,
1997), the All-sky Compiled Catalogue of 2.5 million stars (ASCC-2.5,
2001), and the Tycho-2 Spectral Type Catalog (2003).  CJL acknowledges
support from NASA ORIGINS Grant NAG 5-13041.

\bibliographystyle{aa} 
\bibliography{../dark-refs}

\begin{thebibliography}{26}
\expandafter\ifx\csname natexlab\endcsname\relax\def\natexlab#1{#1}\fi

\bibitem[{{Alves} \& {Franco}(2006)}]{2006MNRAS.366..238A}
{Alves}, F.~O. \& {Franco}, G.~A.~P. 2006, \mnras, 366, 238

\bibitem[{{Alves} {et~al.}(2007){Alves}, {Lombardi}, \&
  {Lada}}]{2007A&A...462L..17A}
{Alves}, J., {Lombardi}, M., \& {Lada}, C.~J. 2007, \aap, 462, L17

\bibitem[{{Bertout} {et~al.}(1999){Bertout}, {Robichon}, \&
  {Arenou}}]{1999A&A...352..574B}
{Bertout}, C., {Robichon}, N., \& {Arenou}, F. 1999, \aap, 352, 574

\bibitem[{{Chini}(1981)}]{1981A&A....99..346C}
{Chini}, R. 1981, \aap, 99, 346

\bibitem[{{de Zeeuw} {et~al.}(1999){de Zeeuw}, {Hoogerwerf}, {de Bruijne},
  {Brown}, \& {Blaauw}}]{1999AJ....117..354D}
{de Zeeuw}, P.~T., {Hoogerwerf}, R., {de Bruijne}, J.~H.~J., {Brown}, A.~G.~A.,
  \& {Blaauw}, A. 1999, \aj, 117, 354

\bibitem[{{Greene} {et~al.}(1994){Greene}, {Wilking}, {Andre}, {Young}, \&
  {Lada}}]{1994ApJ...434..614G}
{Greene}, T.~P., {Wilking}, B.~A., {Andre}, P., {Young}, E.~T., \& {Lada},
  C.~J. 1994, \apj, 434, 614

\bibitem[{{Hughes} {et~al.}(1993){Hughes}, {Hartigan}, \&
  {Clampitt}}]{1993AJ....105..571H}
{Hughes}, J., {Hartigan}, P., \& {Clampitt}, L. 1993, \aj, 105, 571

\bibitem[{{Humphreys}(1978)}]{1978ApJS...38..309H}
{Humphreys}, R.~M. 1978, \apjs, 38, 309

\bibitem[{{Kharchenko}(2001)}]{2001KFNT...17..409K}
{Kharchenko}, N.~V. 2001, Kinematika i Fizika Nebesnykh Tel, 17, 409

\bibitem[{{Knude} \& {Hog}(1998)}]{1998A&A...338..897K}
{Knude}, J. \& {Hog}, E. 1998, \aap, 338, 897

\bibitem[{{Knude} \& {Nielsen}(2001)}]{2001A&A...373..714K}
{Knude}, J. \& {Nielsen}, A.~S. 2001, \aap, 373, 714

\bibitem[{{Krauuter}(1991)}]{1991ESOSR..11....1R}
{Krauuter}, J. 1991, in European Southern Observatory Scientific Report,
  Vol.~11, {Low Mass Star Formation in Southern Molecular Clouds.}, ed.
  B.~{Reipurth}, J.~{Brand}, J.~G.~A. {Wouterloot}, G.~H. {Herbig},
  B.~{Pettersson}, R.~D. {Schwartz}, L.-A. {Nyman}, J.~{Krautter}, J.~A.
  {Graham}, B.~A. {Wilking}, \& C.~{Eiroa}, 127

\bibitem[{{Landolt-B\"ornstein}(1982)}]{LB}
{Landolt-B\"ornstein}. 1982, Numerical Data and Functional Relationships in
  Science and Technology, Vol.~2B, Stars and star clusters (Berlin:
  Springer-Verlag), 15

\bibitem[{{Loinard} {et~al.}(2008){Loinard}, {Torres}, {Mioduszewski}, \&
  {Rodr\'iguez}}]{Loinard}
{Loinard}, L., {Torres}, R.~M., {Mioduszewski}, A.~J., \& {Rodr\'iguez}, L.~F.
  2008, in Proceedings of the IAU Symposium 248

\bibitem[{{Lombardi} \& {Alves}(2001)}]{2001A&A...377.1023L}
{Lombardi}, M. \& {Alves}, J. 2001, \aap, 377, 1023

\bibitem[{{Lombardi} {et~al.}(2006){Lombardi}, {Alves}, \&
  {Lada}}]{2006A&A...454..781L}
{Lombardi}, M., {Alves}, J., \& {Lada}, C.~J. 2006, \aap, 454, 781

\bibitem[{{Lombardi} {et~al.}(2008){Lombardi}, {Alves}, \&
  {Lada}}]{Lombardi07a}
{Lombardi}, M., {Alves}, J., \& {Lada}, C.~J. 2008, submitted to \aap

\bibitem[{{Mamajek}(2007)}]{2007arXiv0709.0505M}
{Mamajek}, E.~E. 2007, ArXiv e-prints, 709

\bibitem[{{Perryman} {et~al.}(1997){Perryman}, {Lindegren}, {Kovalevsky},
  {Hoeg}, {Bastian}, {Bernacca}, {Cr{\' e}z{\' e}}, {Donati}, {Grenon}, {van
  Leeuwen}, {van der Marel}, {Mignard}, {Murray}, {Le Poole}, {Schrijver},
  {Turon}, {Arenou}, {Froeschl{\' e}}, \& {Petersen}}]{1997A&A...323L..49P}
{Perryman}, M.~A.~C., {Lindegren}, L., {Kovalevsky}, J., {et~al.} 1997, \aap,
  323, L49

\bibitem[{{Rebull} {et~al.}(2004){Rebull}, {Wolff}, \&
  {Strom}}]{2004AJ....127.1029R}
{Rebull}, L.~M., {Wolff}, S.~C., \& {Strom}, S.~E. 2004, \aj, 127, 1029

\bibitem[{{Rieke} \& {Lebofsky}(1985)}]{1985ApJ...288..618R}
{Rieke}, G.~H. \& {Lebofsky}, M.~J. 1985, \apj, 288, 618

\bibitem[{{Tanner}(1991)}]{Tanner}
{Tanner}, M. 1991, Tools for Statistical Inference, lecture notes in statistics
  edn., Vol.~67 (Berlin: Springer-Verlag)

\bibitem[{{Wichmann} {et~al.}(1998){Wichmann}, {Bastian}, {Krautter},
  {Jankovics}, \& {Rucinski}}]{1998MNRAS.301L..39W}
{Wichmann}, R., {Bastian}, U., {Krautter}, J., {Jankovics}, I., \& {Rucinski},
  S.~M. 1998, \mnras, 301, L39+

\bibitem[{{Wilking} {et~al.}(1989){Wilking}, {Lada}, \&
  {Young}}]{1989ApJ...340..823W}
{Wilking}, B.~A., {Lada}, C.~J., \& {Young}, E.~T. 1989, \apj, 340, 823

\bibitem[{{Wilson}(2002)}]{WilsonPhD}
{Wilson}, B.~A. 2002, PhD thesis, University of Bristol

\bibitem[{{Wright} {et~al.}(2003){Wright}, {Egan}, {Kraemer}, \&
  {Price}}]{2003AJ....125..359W}
{Wright}, C.~O., {Egan}, M.~P., {Kraemer}, K.~E., \& {Price}, S.~D. 2003, \aj,
  125, 359

\end{thebibliography}

\end{document}